# *TDSelector*: A Training Data Selection Method for Cross-Project Defect Prediction

Peng He[a], Yutao Ma[b,d,*], Bing Li[c,d]

[a]*School of Computer Science and Information Engineering, Hubei University, Wuhan 430062, China*
[b]*State Key Laboratory of Software Engineering, Wuhan University, Wuhan 430072, China*
[c]*International School of Software, Wuhan University, Wuhan 430079, China*
[d]*Research Center for Complex Network, Wuhan University, Wuhan 430072, China*

**Abstract**: *Context*: In recent years, cross-project defect prediction (CPDP) attracted much attention and has been validated as a feasible way to address the problem of local data sparsity in newly created or inactive software projects. Unfortunately, the performance of CPDP is usually poor, and low quality training data selection has been regarded as a major obstacle to achieving better prediction results. To the best of our knowledge, most of existing approaches related to this topic are only based on instance similarity. *Objective*: The objective of this work is to propose an improved training data selection method for CPDP that considers both similarity and the number of defects each training instance has (denoted by *defects*), which is referred to as *TDSelector*, and to demonstrate the effectiveness of the proposed method. *Method*: First, *TDSelector* is constructed in terms of a linear weighted function of similarity and *defects*. Second, the basic defect predictor used in our experiments was built by using the Logistic Regression classification algorithm. Third, we analyzed the impacts of different combinations of similarity and the normalization of *defects* on prediction performance, and then compared our method (with the best combination) and two competing methods using statistical methods. *Results*: Our experiments were conducted on 14 projects (including 15 data sets) collected from two public repositories. The results indicate that, in a specific CPDP scenario, the *TDSelector*-based bug predictor performs, on average, better than those based on the baseline methods, and the AUC (area under ROC curve) values are increased by up to 10.6 and 4.3%, respectively. Besides, an additional experiment shows that selecting those instances with more bugs directly as training data can further improve the performance of the bug predictor trained by our method. *Conclusion*: The findings suggest that (1) the inclusion of *defects* is indeed helpful to select high quality training instances for CPDP, so as to improve prediction performance, and (2) the combination of Euclidean distance and Linear normalization is the preferred way for *TDSelector*.

# 1 Introduction

The past decades have witnessed the development of software defect prediction, which is now one of the most active research topics in the field of Software Engineering. Due to the lack of training data available on the Internet, most early studies usually trained predictors (also known as prediction models) from the historical data on software defects/bugs in the same software project and predicted defects in its upcoming release versions [1]. This type of approaches is referred to as Within-Project Defect Prediction (WPDP). However, WPDP has an obvious drawback when a newly-created or inactive project has little historical data on defects.

To address the above issue, researchers in this field have attempted to apply defect predictors built for one project to other projects [2-4]. This type of methods is termed as Cross-Project Defect Prediction (CPDP). The main purpose of CPDP is to predict defect-prone instances (such as classes) in a project based on the defect data collected from other projects on those public software repositories like PROMISE[1]. The feasibility and potential usefulness of cross-project predictors built with a number of software metrics have been validated [1, 3, 5, 6], but how to improve the performance of CPDP models is still an open issue.

Peter *et al.* [5] argued that selecting appropriate training data from a software repository became a major issue for CPDP. Moreover, some researchers also suggested that the success rate of CPDP models could be drastically improved when using a suitable training data set [1, 7]. That is to say, the selection of training data of quality could be a key breakthrough on the above issue. On the other hand, it is no doubt that labeled defect data available on the Internet will continue to grow smartly. Thus, the construction of an appropriate training data set gathered from a large number of projects on public software repositories is indeed a challenge for CPDP [7].

As far as we know, although previous studies on CPDP have taken different types of software metrics into account during the process of selecting relevant training samples, none of them considered the number of defects contained in each sample (denoted by *defects*). But in fact, we argue that it is also an important factor to consider. Fortunately, some studies have empirically demonstrated the relevance of *defects* to defect prediction. For example, "modules with faults in the past are likely to have faults in the future" [8], "17% to 54% of the high-fault files of release *i* are still high-fault in release *i*+1" [9], "cover 73%-95% of faults by selecting 10% of the most fault prone source code file" [10], and "the number of defects found in the previous release of file correlates with its current defect count on a high level" [11].

Does the selection of training data considering *defects* improve the prediction performance of CPDP models? If the answer is "Yes", on the one hand, it is helpful to further validate the feasibility of CPDP; on the other hand, it will contribute to better software defect predictors by making full use of those defect data sets available on the Internet.

The objective of our work is to propose an improved method of training data selection for CPDP by introducing the information of *defects*. Unlike the prior studies similar to our work, such as [12] and [5], which focus mainly on the similarity between instances from training set and test set, this paper gives a comprehensive account of two factors, namely similarity and *defects*. Moreover, the proposed method, called *TDSelector*, can automatically optimize their weights to achieve the best result. In brief, our main contributions to the current state of research on CPDP are summarized as follows.

(1) Considering both similarity and *defects*, we proposed a simple and easy-to-use training data selection method for CPDP (i.e. *TDSelector*), which is based on an improved scoring scheme that rank all possible training instances. In particular, we designed an algorithm to calculate their weights automatically, so as to obtain the best prediction result.

---

[1]http://openscience.us/repo/

(2) To validate the effectiveness of our method, we conducted an elaborate empirical study based on 15 data sets collected from PROMISE and AEEEM[2], and the experimental results show that, in a specific CPDP scenario (i.e. *many-to-one* [13]), the *TDSelector*-based defect predictor outperforms its rivals that were built with two competing methods in terms of prediction precision.

With these technical contributions, our study could complement previous work on CPDP with respect to training data selection. In particular, we provide a reasonable scoring scheme as well as a more comprehensive guideline for developers to choose appropriate training data to train a defect predictor in practice.

The rest of this paper is organized as follows. In Section 2, we reviewed the related work of this topic; Section 3 presents the preliminaries to our work; Section 4 describes the proposed method *TDSelector*, Section 5 introduces our experimental setup, and Section 6 shows the primary experimental results; a detailed discussion of some issues including potential threats to the validity of our study is presented in Section 7; in the end, Section 8 summaries this paper and presents our future work.

## 2    Related Work

### 2.1    Cross-Project Defect Prediction

To the best of our knowledge, the earliest study on CPDP was performed by Briand *et al.* [2], who applied the models built on an open-source project Xpose to another open-source project Jwriter. Several years later, Zimmermann *et al.* [4] conducted a large-scale experiment on data vs. domain vs. process and found that only 3.4% of 622 cross-project predictions actually worked. In particular, they also found that CPDP was not symmetrical between Firefox and Microsoft IE. In other words, the result showed that Firefox was a sound defect predictor for Microsoft IE, but not vice versa.

Many studies were carried out to validate the feasibility of CPDP in the last five years. For example, Turhan *et al.* [12] proposed a cross-company defect prediction approach using defect data from other companies to build predictors for target projects. They found that the proposed method increased the probability of defect detection at the cost of increasing false positive rate. Rahman *et al.* [6] evaluated the validity of CPDP in term of cost-effectiveness and confirmed that, for the collected nine defect data sets, the prediction performance of CPDP is comparable to that of WPDP. Furthermore, He *et al.* [1] conducted three experiments on 34 data sets from PROMISE, and they concluded that training data from other projects, compared with those from the same project, could provide similar or better prediction results.

A growing number of researchers in this field have attached great importance to improving the performance of CPDP. For example, Fukushima *et al.* [14] took action to deal with this issue from the perspective of Just-in-Time (JIT) defect prediction, and they verified that JIT CPDP was a viable solution for those projects with little historical defect data. He *et al.* [15] compared the performance between CPDP and WPDP using feature selection techniques. The results indicated that for reduced training data WPDP obtained higher precision, but CPDP in turn achieved a better recall or F-measure. Ma *et al.* [3] utilized the transfer learning method to build faster and highly effective prediction models called Transfer Naive Bayes (TNB). The experimental result showed that TNB was more accurate and less time-consuming than benchmark methods. Inspired by the idea of ensemble learning, some researchers have also studied the performance of CPDP based on ensemble classifiers, and then validated their effects on this issue [16, 17].

---
[2]http://bug.inf.usi.ch

Liu *et al.* [18] studied 17 different machine learning algorithms in the context of CPDP, and they suggested building cross-project prediction models via a Validation-and-Voting learning mechanism. Canfora *et al.* [19] proposed a multi-objective approach for CPDP by taking into account the tradeoff between effectiveness and inspection cost. The results indicated that the proposed multi-objective approach performed better than the single-objective ones, especially in the context of CPDP. Panichella *et al.* [20] also investigated the equivalence of different machine learning techniques in the context of CPDP, and they proposed a combined approach to improve the prediction accuracy of CPDP.

Recently, Zhang *et al.* [21] proposed a universal CPDP model, which was built using a large number of projects collected from SourceForge[3] and Google Code[4]. Their experimental results showed that it was indeed comparable to WPDP. Furthermore, CPDP has been validated to be feasible for different projects which have heterogeneous metric sets. He *et al.* [22] first proposed a CPDP-IFS approach based on the distribution characteristics of both source and target projects to overcome this problem. Nam *et al.* [23] then proposed an improved method HDP, where metric selection and metric matching were introduced to build a defect predictor. Their empirical study on 28 projects showed that about 68% of predictions using the proposed approach outperformed or were comparable to WPDP with statistical significance. Jing *et al.* [24] proposed a unified metric representation UMR for heterogeneous defect data. The experiments on 14 public heterogeneous data sets from four different companies indicated that the proposed approach was more effective to address the problem.

## 2.2 Training Data Selection for CPDP

As mentioned in [5, 25], a fundamental issue for CPDP is selecting the most appropriate training data for building quality defect predictors. Watanabe *et al.* [26] made an early attempt to address the issue, and they proposed a metric compensation-based approach to transform data sets by using the average metric values. In the case of similar domain and size, they argued that it was possible to reuse prediction models between projects written in different programming languages.

Turhan *et al.* [12] used a nearest neighbor filter to select similar data from source projects as training data. According to the experimental results, they concluded that the nearest neighbor filtering technique could filter out the irrelevances and reduce false positive rate when using cross-company data. An improved strategy was then proposed by Peter *et al.* [5]. They designed a filter that used training instances to guide the selection of training data, because training data sets usually contain more information about defects than test data. The experimental results showed that defect predictors built with the proposed filter and cross-company data performed better.

He *et al.* [1] presented an approach to automatically selecting suitable training data for CPDP. More specifically, they calculated the similarity between training data and test data using distributional characteristics, such as the mode, mean, median, maximum, minimum and so on, and then picked out the relevant data. Similarly, Herbold proposed two distance-based strategies for selecting appropriate training data based on distributional characteristics [7]. The author evaluated his strategies with a case study of 14 open-source projects, and the results demonstrated that the proposed methods could improve the achieved success rate significantly, although they were still unable to compete with WPDP. He *et al.* [27] discussed this problem in detail from the perspective of data granularity, i.e. release level and instance level. They presented a two-step method for training data selection. The results indicated that the predictor built based on Naive Bayes could achieve fairly good performance when using the method together with Peter filter [5].

---

[3] https://sourceforge.net/
[4] http://code.google.com/

With regard to the data imbalance problem of defect data sets, Ryu *et al.* [28] recently proposed a method of hybrid instance selection using nearest neighbor (HISNN). The approach HISNN utilized the outlier removing technique and the nearest neighbor approach to filter out instances in a source project that might hinder prediction performance. Finally, instances in the target project were classified by the Naive Bayes algorithm using the selected training data. Their results suggested that those instances which had strong local knowledge could be identified via nearest-neighbors with the same class label.

The above-mentioned existing studies aimed at reducing the gap in prediction performance between WPDP and CPDP. Although they are making progress towards the goal, there is clearly lots of room for improvement. For this reason, in this paper we proposed a selection approach to training data based on an improved strategy for instance ranking instead of a single strategy for similarity calculation, which was used in many prior studies [1, 5, 7, 12, 27].

## 3 Preliminaries

In our context, a defect data set $S$ contains $m$ instances, which is represented as $S = \{I_1, I_2, ..., I_m\}$. Instance $I_i$ is an object class represented as $I_i = \{f_{i1}, f_{i2}, ..., f_{in}\}$, where $f_{ij}$ is the $j^{th}$ metric value of instance $I_i$ and $n$ is the number of metrics (also known as features). Given a source data set $S_s$ and a target data set $S_t$, CPDP aims to perform a prediction in $S_t$ using the knowledge extracted from $S_s$, where $S_s \neq S_t$ (see Fig.1a). In this paper source and target data sets have the same set of metrics, and they may differ in distributional characteristics of metric values.

To improve the performance of CPDP, several strategies used to select appropriate training data have been put forward (see Fig.1b), e.g. Turhan *et al.* [12] filtered out those irrelevant training instances by returning $k$-nearest neighbors for each test instance.

### 3.1 An Example of Training Data Selection

First of all, we introduce a typical method for training data selection at the instance level, and a simple example is used to illustrate this method. Note that the selection strategy for other levels of data granularity such as release refers to [7].

Fig.2 shows a training set $S_s$ (including five instances) and a test set $S_t$ (including an instance). Here each instance contains five metrics and a classification label (i.e. 0 or 1). An instance is defect-free (label = 0) only if its *defects* is equal to 0; otherwise, it is defective (label = 1). According to the $k$ nearest-neighbor method based on Euclidean distance, we can rank all the five training instances in terms of their distances from the test instance. Due to the same nearest distance from test instance $I_{test}$, it is clear that three instances $I_1$, $I_2$ and $I_5$ are suitable for use as training instances when $k$ is set to 1. For the three instances, $I_2$ and $I_5$ have the same metric values, but $I_2$ is labeled as a defective instance because it contains a bug. In this case, $I_1$ will be selected with the same probability as that of $I_2$, regardless of the number of defects they includes.

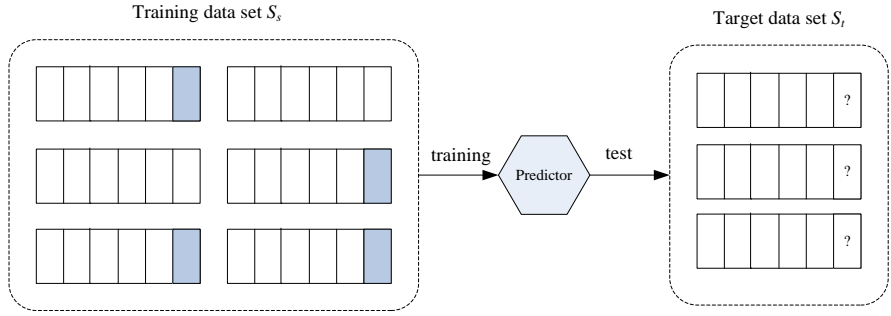

(a) General CPDP

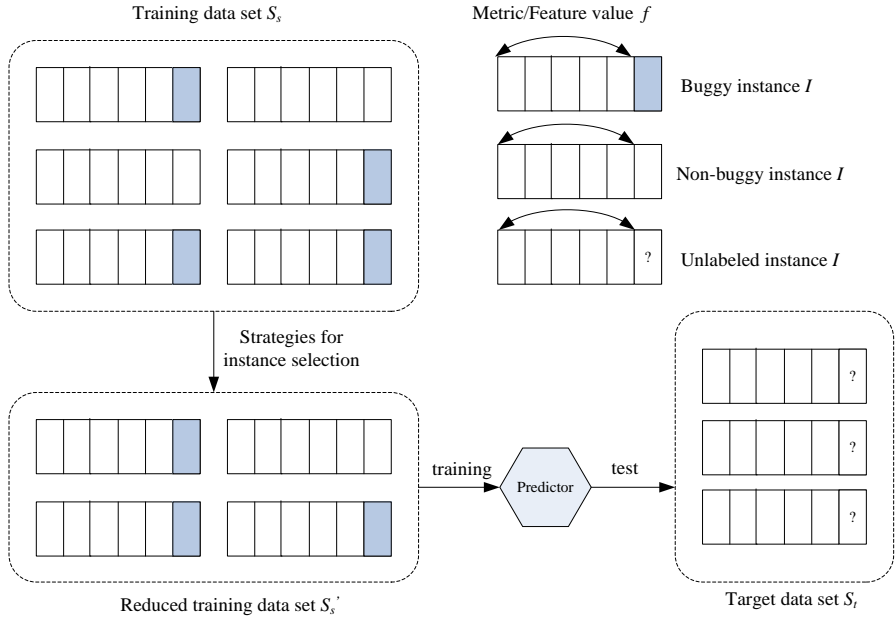

(b) Improved CPDP using training data selection

**Figure 1.** Two CPDP scenarios used in this paper.

In this way, those instances most relevant to the test one can be quickly determined. Clearly, the goal of training data selection is to preserve the representative training instances in $S_s$ as much as possible.

|  |  | $f_1$ | $f_2$ | $f_3$ | $f_4$ | Label (defects) |
|---|---|---|---|---|---|---|
| $S_s$ | $I_1$ | 0.1 | 0 | 0.5 | 0 | 1(3) |
|  | $I_2$ | 0.1 | 0 | 0 | 0.5 | 1(1) |
|  | $I_3$ | 0.4 | 0.3 | 0 | 0.1 | 0 |
|  | $I_4$ | 0 | 0 | 0.4 | 0 | 0 |
|  | $I_5$ | 0.1 | 0 | 0 | 0.5 | 0 |
| $S_t$ | $I_{test}$ | 0.1 | 0 | 0.5 | 0.5 | ? |

distance($I_i, I_{test}$)

| rank | instance |
|---|---|
| 1 | $I_1, I_2, I_5$ |
| 2 | $I_4$ |
| 3 | $I_3$ |

**Figure 2.** An example of the selection of training instances.

### 3.2 General Process of Training Data Selection

Before presenting our approach, we describe a general training data selection process, which consists of three main steps: TDS (training data set) setup, ranking, and duplicate removal.

**TDS setup**: For each target project with little historical data, we need to set up an initial TDS where training data are collected from other projects. To simulate this scenario of CPDP, in this paper any defect data from the target project must be excluded from the initial TDS. Note that different release versions of a project actually belong to the same project. A simple example is visualized in Fig.3.

**Ranking**: Once the initial TDS is determined, an instance will be treated as a metric vector $I$, as mentioned above. For each test instance, one can calculate its relevance to each training instance, and then ranks these training instances in terms of their similarity based on software metrics. Note that a wide variety of software metrics, such as source code metrics, process metrics, previous defects, and code churn, have been used as features for CPDP approaches to improve their prediction performance.

**Duplicate removal**: Let $l$ be the size of test set. For each test instance, if we select its $k$ nearest neighbors from the initial TDS, there are a total of $k \times l$ candidate training instances. Considering that these selected instances may not be unique (i.e. a training instance can be the nearest neighbor of multiple test instances), after removing the duplicate ones, they form the final training set which is a subset of the initial TDS.

## 4 Our Approach *TDSelector*

To improve the prediction performance of CPDP, we leverage the following observations.

**Similar instances**: Given a test instance, we can examine its similar training instances that were labeled before. The defect-proneness shared by similar training instances could help us identify the probability that a test instance is defective. Intuitively, two instances are more likely to have the same state if their metric values are very similar.

**Number of defects (*defects*)**: During the selection process, when several training instances have the same distance from a test instance, we need to determine which one should be ranked higher. According to our experiences in software defect prediction and other researchers' studies on the quantitative analysis of previous defect prediction approaches [29, 30], we believe that more attention should be paid to those training instances with more defects in practice.

The selection of training data based on instance similarity has been used in some prior studies [5, 12, 31]. However, to the best of our knowledge, the information about *defects* has not been fully utilized. So, in this paper we attempt to propose a training data selection approach combing such information and instance similarity.

### 4.1 Overall Structure of *TDSelector*

Fig.3 shows the overall structure of the proposed approach to training data selection named *TDSelector*. Before selecting appropriate training data for CPDP, we have to set up a test set and its corresponding initial TDS. For a given project treated as the test set, all the other projects (except the target project) available at hand are used as the initial TDS. This is the so-called *many-to-one* (M2O) scenario for CPDP [13]. It is quite different from the typical O2O (*one-to-one*) scenario, where only one randomly selected project is treated as the training set for a given target project (viz. test set).

When both of the sets are given, the ranks of training instances are calculated based on the similarity of software metrics and then returned for each test instance. For the initial TDS, we also collect each training instance's *defects*, and thus rank these instances by their *defects*. Then, we rate each training instance by combining the two types of ranks in some way, and identify the top-*k* training instances for each test instance according to their final scores. Finally, we use the predictor trained with the final TDS to predict defect proneness in the test set. We describe the core component of *TDSelector*, namely scoring scheme, in the following subsection.

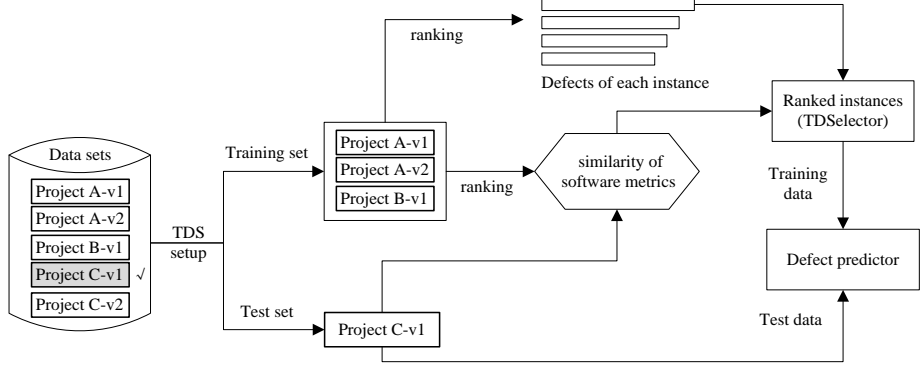

**Figure 3.** The overall structure of *TDSelector* for CPDP.

## 4.2 Scoring Scheme

**Table 1.** Similarity indexes and normalization methods used in this paper.

| | | |
|---|---|---|
| Similarity | Cosine | $\cos(X,Y) = \dfrac{\sum_{k=1}^{n} x_k y_k}{\sqrt{\sum_{k=1}^{n} x_k^2}\sqrt{\sum_{k=1}^{n} y_k^2}}$ |
| | Euclidean distance | $d(X,Y) = \sqrt{\sum_{k=1}^{n}(x_k - y_k)^2}$ |
| | Manhattan distance | $d(X,Y) = \sum_{k=1}^{n}|x_k - y_k|$ |
| Normalization | Linear | $N(x) = \dfrac{x - x_{\min}}{x_{\max} - x_{\min}}$ |
| | Logistic | $N(x) = \dfrac{1}{1+e^{-x}} - 0.5$ |
| | Square root | $N(x) = 1 - \dfrac{1}{\sqrt{1+x}}$ |
| | Logarithmic | $N(x) = \log_{10}(x+1)$ |
| | Inverse cotangent | $N(x) = \dfrac{\arctan(x) * 2}{\pi}$ |

For each instance in training set and test set which is treated as a vector of features (viz. software metrics), we calculate the similarity between them in terms of similarity index (such as

Cosine similarity, Euclidean distance, and Manhattan distance, as shown in Table 1). Training instances are then ranked by the similarity between each of them and a given test instance.

For instance, the cosine similarity between a training instance $I_p$ and the target instance $I_q$ is computed via their vector representations, described as follows.

$$Sim(I_p, I_q) = \frac{\vec{I_p} \cdot \vec{I_q}}{\|I_p\| \times \|I_q\|} = \frac{\sum_{i=1}^{n}(f_{pi} \times f_{qi})}{\sqrt{\sum_{i=1}^{n} f_{pi}^2} \times \sqrt{\sum_{i=1}^{n} f_{qi}^2}}, \quad (1)$$

where $\vec{I_p}$ and $\vec{I_q}$ are the metric vectors for $I_p$ and $I_q$, respectively, and $f_{*i}$ represents the $i^{th}$ metric value of instance $I_*$.

Additionally, for each training instance, we also consider the factor *defects* in order to further enrich the ranking of its relevant instances. The assumption here is that the more the previous *defects*, the richer the information of an instance. So, we propose a scoring scheme to rank those candidate training instances, defined as below.

$$Score(I_p, I_q) = \alpha Sim(I_p, I_q) + (1-\alpha) N(defect_p), \quad (2)$$

where $defect_p$ represents the *defects* of $I_p$, $\alpha$ is a weighting factor ($0 \leq \alpha \leq 1$) which is learned from training data using Algorithm 1 (see Fig.4), and $N(defect_p)$ is a function used to normalize *defects* with values ranging from 0 to 1.

---

**Algorithm 1** Optimizing the parameter $\alpha$

**Input:**
1. Candidate TDS $S_s = \{I_{s1}, I_{s2}, \cdots, I_{sm}\}$, test set $S_t = \{I_{t1}, I_{t2}, \cdots, I_{tl}\}$ ($m > l$),
2. $defect_s = \{defect(I_{s1}), defect(I_{s2}), \cdots, defect(I_{sm})\}$, and $k = 10$

**Output:**
3. $\alpha$ ($\alpha \in [0,1]$)

**Method:**
4.   Initialize $\alpha = 0$, $S_s(\alpha) = \emptyset$;
5.   **While** ($\alpha \leq 1$) **do**
6.     **For** $i = 1; i \leq l; i++$
7.       **For** $j = 1; j \leq m; j++$
8.         $Score(I_{ti}, I_{sj}) = \alpha\, Sim(I_{ti}, I_{sj}) + (1-\alpha)\, N(defect(I_{sj}))$;
9.       **End For**
10.      $descSort(\{Score(I_{ti}, I_{sj}) | j = 1..m\})$; //sort $m$ training instances in descending order
11.      $S_s(\alpha) = S_s(\alpha) \cup \{\text{Top-}k \text{ training instances}\}$; //select the top $k$ instances
12.    **End For**
13.    $AUC \Leftarrow S_s(\alpha) \xrightarrow{CPDP} S_t$; //prediction result
14.    $\alpha = \alpha + 0.1$;
15.  **End While**
16.  **Return** ($\alpha | \max_\alpha AUC$);

---

**Figure 4.** Algorithm of parameter optimization

Normalization is a commonly-used data preprocessing technique for mathematics and computer science [32]. Graf *et al.* [33] have confirmed that normalization can improve prediction performance of classification models. For this reason, we normalize the *defects* of training instances when using *TDSelector*. As you know, there are many normalization methods. In this study, we introduce five typical normalization methods used in machine learning [32, 34]. The description and formulas of the five normalization methods are listed in Table 1.

For each test instance, the top-*k* training instances ranked in terms of their scores will be returned. Hence, the final TDS is composed by merging the sets of the top-*k* training instances for each test instance when those duplicate instances are removed.

## 5 Experimental Setup

### 5.1 Research Questions

Our experiments were conducted to find empirical evidences that answer the following three research questions.

*RQ1: Does the consideration of defects improve the performance of CPDP?*

Unlike the previous methods [1, 5, 7, 12, 27], *TDselector* ranks candidate training instances in terms of both *defects* and metric-based similarity. To evaluate the effectiveness of the proposed method considering the additional information of *defects*, we tested *TDSelector* according to the experimental data described in Subsection 5.2. According to Eq. (2), we also empirically analyzed the impact of the parameter $\alpha$ on prediction results.

*RQ2: Which combination of similarity and normalization is more suitable for TDSelector?*

Eq. (2) is comprised of two parts, namely similarity and the normalization of *defects*. For each part, several commonly-used methods can be adopted in our context. To fully take advantage of *TDSelector*, one would wonder which combination of similarity and normalization should be chosen. Therefore, it is necessary to compare the effects of different combinations of similarity and normalization methods on prediction results, and to determine the best one for *TDSelector*.

*RQ3: Can TDSelector-based CPDP outperform the baseline methods?*

Cross-project prediction has attracted much research interest in recent years, and a few CPDP approaches using training data selection has also been proposed, e.g. Peter filter based CPDP [5] (labeled as baseline1) and TCA+ (Transfer Component Analysis) based CPDP [35] (labeled as baseline2). To answer the third question, we compared *TDSelector*-based CPDP proposed in this paper with the above two state-of-the-art methods.

### 5.2 Data Collection

To evaluate the effectiveness of *TDSelector*, in this paper we used 14 open-source projects written in Java on two on-line public software repositories, namely PROMISE [36] and AEEEM [29]. The data statistics of the 14 projects in question are presented in Table 2, where #*Instance* and #*Defect* are the numbers of instances and defective instances, respectively, and %*Defect* is the proportion of defective instances to the total number of instances. Each instance in these projects represents a file of object class and consists of two parts, namely software metrics and *defects*.

The first repository, PROMISE, was collected by Jureczko and Spinellis [36]. The information of defects and 20 source code metrics for the projects on PROMISE have been validated and used in several previous studies [1, 7, 12, 27]. The second repository, AEEEM, was collected by D' Ambros *et al.* [29], and each project on it have 76 metrics, including 17 source code metrics, 15 change metrics, 5 previous defect metrics, 5 entropy-of-change metrics, 17 entropy-of-source-code metrics, and 17 churn-of-source-code metrics. AEEEM has been successfully used in [23, 35].

Table 2. Data statistics of the projects used in our experiments.

| Repository | Project | Version | #Instance | #Defect | %Defect |
|---|---|---|---|---|---|
| PROMISE | Ant | 1.7 | 745 | 166 | 22.3% |
| | Camel | 1.6 | 965 | 188 | 19.5% |
| | Ivy | 2.0 | 352 | 40 | 11.4% |
| | Jedit | 3.2 | 272 | 90 | 33.1% |
| | Lucene | 2.4 | 340 | 203 | 59.7% |
| | Poi | 3.0 | 442 | 281 | 63.6% |
| | Synapse | 1.2 | 256 | 86 | 33.6% |
| | Velocity | 1.4 | 196 | 147 | 75.0% |
| | Xalan | 2.6 | 885 | 411 | 46.4% |
| | Xerces | 1.4 | 588 | 437 | 74.3% |
| AEEEM | Equinox | 1.1.2005-6.25.2008 | 324 | 129 | 39.8% |
| | Eclipse JDT core(Eclipse) | 1.1.2005-6.17.2008 | 997 | 206 | 20.7% |
| | Apache Lucene (Lucene2) | 1.1.2005-10.8.2008 | 692 | 20 | 2.9% |
| | Mylyn | 1.17.2005-3.17.2009 | 1,862 | 245 | 13.2% |
| | Eclipse PDE UI (Pde) | 1.1.2005-9.11.2008 | 1,497 | 209 | 14.0% |

### 5.3 Experiment Design

To answer the three research questions, our experimental procedure, which is designed under the context of M2O in the CPDP scenario, is described as follows.

First, as with many prior studies [1, 5, 15, 31], all software metric values in training and test sets were normalized by using the *Z*-score method, because these metrics are different in the scales of numerical values. For the 14 projects on AEEEM and PROMISE, their numbers of software metrics are different. So, the training set for a given test set was selected from the same repository.

Second, to examine whether the consideration of *defects* improve the performance of CPDP, we compared our approach *TDSelector* with NoD, which is a baseline method considering only the similarity between instances, i.e. $\alpha = 1$ in Eq. (2). Since there are three similarity computation methods used in this paper, we designed three different TDSelectors and their corresponding baseline methods based on similarity indexes. The prediction results of each method in question for the 15 test sets were analyzed in terms of mean value and standard deviation. More specifically, we also used the Cliff's delta (*d*) [37], which is a non-parametric effect size measure of how often the values in one distribution are larger than the values in a second distribution, to compare the results generated through our approach and its corresponding baseline method.

Third, according to the results of the second step of this procedure, 15 combinations based on three typical similarity methods for software metrics and five commonly used normalization functions for *defects* were examined by the pairwise comparison method. We then determined which combination is more suitable for our approach according to mean, standard deviation and the Cliff's delta effect size.

Fourth, to further validate the effectiveness of the *TDSelector*-based CPDP predictor, we conducted cross-project predictions for all the 15 test sets using *TDSelector* and two competing methods (i.e. baseline1 and baseline2 introduced in Subsection 5.1). Note that the TDSelector used in this experiment was built with the best combination of similarity and normalization.

After this process is completed, we will discuss the answers to the three research questions of our study.

### 5.4 Classifier and Evaluation Measure

As an underlying machine learning classifier for CPDP, Logistic Regression (LR), which was widely used in many defect prediction literatures [4, 23, 35, 38-41], is also used in this study. All LR classifiers were implemented with Weka[5]. For our experiments, we used the default parameter setting for LR specified in Weka unless otherwise specified.

To evaluate the prediction performance of different methods, in this paper we utilized the area under a Receiver Operating Characteristic curve (AUC). AUC is equal to the probability that a classifier will identify a randomly chosen defective class higher than a randomly chosen defect-free one [42], known as a useful measure for comparing different models. Compared with traditional accuracy measures, AUC is commonly used because it is unaffected by class imbalance and independent from the prediction threshold that is used to decide whether an instance should be classified as a negative instance [6, 43, 44]. The AUC value of 0.5 indicates the performance of a random predictor, and higher AUC values indicate better prediction performance.

## 6 Experimental Results

### 6.1 Answer to RQ1

We compared our approach considering *defects* with the baseline method NoD that selects training data in terms of Cosine similarity. Table 3 shows that, on average, *TDSelector* does achieve an improvement in AUC value across the 15 test sets. Obviously, the average growth rates of AUC value vary from 5.9% to 9.0% when different normalization methods for *defects* were utilized. In addition, all the Cliff's delta (*d*) effect sizes in this table are greater than 0.2, which indicates that each group of 15 prediction results obtained by our approach has a greater effect than that of NoD. In other words, our approach outperforms NoD. In particular, for Jedit, Velocity, Eclipse and Equinox, the improvements of our approach over NoD are substantial. For example, when using the linear normalization method, the AUC values for the four projects are increased by 30.6%, 43.0%, 22.6%, and 39.4%, respectively; moreover, the logistic normalization method for Velocity achieves the biggest improvement in AUC value (viz. 61.7%).

We then compared *TDSelector* with the baseline methods using other widely-used similarity calculation methods, and the results obtained by using Euclidean distance and Manhattan distance to calculate the similarity between instances are presented in Tables 4 and 5. *TDSelector*, compared with the corresponding NoD, achieves the average growth rates of AUC value that vary from 5.9% to 7.7% in Table 4 and from 2.7% to 6.9% in Table 5, respectively. More specifically, the highest growth rate of AUC value in Table 4 is 43.6% for Equinox and in Table 5 is 39.7% for Lucene2. Besides, all the Cliff's delta (*d*) effect sizes in these two tables are also greater than 0.1. Hence, the results indicate that our approach can, on average, improve the performance of those baseline methods without regard to *defects*.

In short, during the process of training data selection, the consideration of *defects* for CPDP can help us to select higher quality training data, thus leading to better classification results.

---
[5]http://www.cs.waikato.ac.nz/ml/weka/

**Table 3.** The best prediction results obtained by the CPDP approach based on *TDSelector* with Cosine similarity. NoD represents the baseline method, + denotes the growth rate of AUC value, the maximum AUC value of different normalization methods is underlined, and each number shown in bold indicates that the corresponding AUC value rises by more than 10%.

| Cosine similarity | | Ant | Xalan | Camel | Ivy | Jedit | Lucene | Poi | Synapse | Velocity | Xerces | Eclipse | Equinox | Lucene2 | Mylyn | Pde | Mean±St.d | d |
|---|---|---|---|---|---|---|---|---|---|---|---|---|---|---|---|---|---|---|
| Linear | α | 0.7 | 0.9 | 0.9 | 1.0 | 0.9 | 1.0 | 0.9 | 1.0 | 0.6 | 0.9 | 0.8 | 0.6 | 0.7 | 0.7 | 0.5 | | |
| | AUC | <u>0.813</u> | <u>0.676</u> | <u>0.603</u> | 0.793 | 0.700 | 0.611 | 0.758 | 0.741 | 0.512 | 0.742 | <u>0.783</u> | <u>0.760</u> | 0.739 | 0.705 | 0.729 | 0.711±0.081 | 0.338 |
| | +(%) | 6.3% | 3.7% | 1.9% | - | **30.6%** | - | 3.0% | - | **43.0%** | 0.3% | **22.6%** | **39.4%** | 4.1% | 5.9% | 4.0% | 9.0% | |
| Logistic | α | 0.7 | 0.5 | 0.7 | 1 | 0.7 | 0.6 | 0.6 | 0.6 | 0.5 | 0.5 | 0 | 0.4 | 0.7 | 0.5 | 0.5 | | |
| | AUC | 0.802 | 0.674 | 0.595 | 0.793 | 0.665 | 0.621 | <u>0.759</u> | <u>0.765</u> | <u>0.579</u> | 0.745 | 0.773 | 0.738 | 0.712 | <u>0.707</u> | 0.740 | 0.711±0.070 | 0.351 |
| | +(%) | 4.8% | 3.4% | 0.5% | - | **24.1%** | 1.6% | 3.1% | 3.2% | **61.7%** | 0.7% | **21.0%** | **35.5%** | 0.3% | 6.2% | 5.6% | 9.0% | |
| Square root | α | 0.7 | 0.7 | 0.6 | 0.6 | 0.7 | 0.6 | 0.7 | 0.9 | 0.5 | 1 | 0.4 | 0.6 | 0.6 | 0.6 | 0.6 | | |
| | AUC | 0.799 | 0.654 | 0.596 | <u>0.807</u> | <u>0.735</u> | <u>0.626</u> | 0.746 | 0.762 | 0.500 | 0.740 | 0.774 | 0.560 | 0.722 | 0.700 | 0.738 | 0.697±0.091 | 0.249 |
| | +(%) | 4.4% | 0.3% | 0.7% | 1.8% | **37.1%** | 2.5% | 1.4% | 2.8% | **39.7%** | - | **21.0%** | 2.8% | 1.7% | 5.3% | 5.3% | 6.9% | |
| Logarithmic | α | 0.6 | 0.6 | 0.9 | 1.0 | 0.7 | 1.0 | 0.7 | 0.7 | 0.5 | 0.9 | 0.5 | 0.5 | 0.6 | 0.6 | 0.6 | | |
| | AUC | 0.798 | 0.662 | 0.594 | 0.793 | 0.731 | 0.611 | 0.748 | 0.744 | 0.500 | 0.758 | 0.774 | 0.700 | <u>0.755</u> | 0.702 | <u>0.741</u> | 0.707±0.083 | 0.351 |
| | +(%) | 4.3% | 1.5% | 0.3% | - | **36.4%** | - | 1.6% | 0.4% | **39.7%** | 2.4% | **21.2%** | **28.5%** | 6.3% | 5.5% | 5.8% | 8.5% | |
| Inverse cotangent | α | 0.7 | 1.0 | 1.0 | 1.0 | 0.7 | 1.0 | 0.7 | 1.0 | 0.6 | 0.7 | 0 | 0.7 | 0.7 | 0.7 | 0.7 | | |
| | AUC | 0.798 | 0.652 | 0.592 | 0.793 | 0.659 | 0.611 | 0.749 | 0.741 | 0.500 | <u>0.764</u> | 0.773 | 0.556 | 0.739 | 0.695 | 0.734 | 0.690±0.092 | 0.213 |
| | +(%) | 4.3% | - | - | - | **22.9%** | - | 1.8% | - | **39.7%** | 3.2% | **21.0%** | 2.1% | 4.1% | 4.4% | 4.8% | 5.9% | |
| NoD (α = 1) | | 0.765 | 0.652 | 0.592 | 0.793 | 0.536 | 0.611 | 0.736 | 0.741 | 0.358 | 0.740 | 0.639 | 0.543 | 0.709 | 0.665 | 0.701 | 0.652±0.113 | |

**Table 4.** The best prediction results obtained by the CPDP approach based on *TDSelector* with Euclidean distance.

| Euclidean distance | | Ant | Xalan | Camel | Ivy | Jedit | Lucene | Poi | Synapse | Velocity | Xerces | Eclipse | Equinox | Lucene2 | Mylyn | Pde | Mean±St.d | d |
|---|---|---|---|---|---|---|---|---|---|---|---|---|---|---|---|---|---|---|
| Linear | α | 0.9 | 0.9 | 1.0 | 0.9 | 0.9 | 0.8 | 1.0 | 1.0 | 0.8 | 0.8 | 0 | 0.6 | 1.0 | 0.8 | 0.8 | | |
| | AUC | 0.795 | 0.727 | 0.598 | 0.826 | <u>0.793</u> | 0.603 | 0.714 | 0.757 | 0.545 | <u>0.775</u> | 0.773 | 0.719 | 0.722 | <u>0.697</u> | 0.744 | 0.719±0.080 | 0.369 |
| | +(%) | 1.3% | 6.8% | - | 0.9% | **32.2%** | 1.9% | - | - | **11.7%** | 5.2% | **17.6%** | **43.0%** | - | 1.1% | 9.6% | 7.7% | |
| Logistic | α | 0.7 | 0.8 | 0.4 | 0.7 | 0.7 | 0.5 | 0.6 | 0.9 | 0.9 | 0.9 | 0 | 0.7 | 1.0 | 1.0 | 0.9 | | |
| | AUC | 0.787 | <u>0.750</u> | <u>0.603</u> | <u>0.832</u> | 0.766 | 0.613 | 0.716 | 0.767 | 0.556 | 0.745 | 0.773 | 0.698 | 0.722 | 0.690 | 0.730 | 0.717±0.075 | 0.360 |
| | +(%) | 0.3% | **10.1%** | 0.8% | 1.6% | **27.7%** | 3.5% | 0.3% | 1.3% | **13.9%** | 1.1% | **17.6%** | **38.8%** | - | - | 7.5% | 7.2% | |
| Square root | α | 0.7 | 0.8 | 1.0 | 0.7 | 0.8 | 0.6 | 0.7 | 0.7 | 0.7 | 1.0 | 0.7 | 0.8 | 1.0 | 1.0 | 0.9 | | |
| | AUC | <u>0.796</u> | 0.743 | 0.598 | 0.820 | 0.720 | 0.618 | <u>0.735</u> | 0.786 | 0.564 | 0.737 | 0.774 | 0.696 | 0.722 | 0.690 | <u>0.750</u> | 0.715±0.076 | 0.342 |
| | +(%) | 1.4% | 9.1% | - | 0.1% | **20.0%** | 4.4% | 2.9% | 3.8% | **15.6%** | - | **17.8%** | **38.4%** | - | - | **10.5%** | 7.0% | |
| Logarithmic | α | 0.7 | 0.8 | 1.0 | 1.0 | 0.8 | 0.6 | 1.0 | 1.0 | 0.9 | 0.9 | 0.9 | 0.8 | 1.0 | 1.0 | 0.9 | | |
| | AUC | 0.794 | 0.746 | 0.598 | 0.819 | 0.722 | 0.607 | 0.714 | 0.757 | <u>0.573</u> | 0.739 | <u>0.778</u> | <u>0.722</u> | 0.722 | 0.690 | 0.748 | 0.715±0.072 | 0.324 |
| | +(%) | 1.1% | 9.5% | - | - | **20.3%** | 2.5% | - | - | **17.4%** | 0.3% | **18.5%** | **43.6%** | - | - | **10.3%** | 7.0% | |
| Inverse cotangent | α | 0.8 | 0.9 | 0.6 | 0.8 | 0.8 | 0.7 | 1.0 | 0.8 | 0.6 | 0.7 | 0 | 0.9 | 0.9 | 1.0 | 0.9 | | |
| | AUC | <u>0.796</u> | 0.749 | <u>0.603</u> | 0.820 | 0.701 | <u>0.623</u> | 0.714 | <u>0.787</u> | 0.538 | 0.750 | 0.773 | 0.589 | <u>0.763</u> | 0.690 | 0.722 | 0.708±0.084 | 0.280 |
| | +(%) | 1.4% | **10.0%** | 0.8% | 0.1% | **16.8%** | 5.2% | - | 4.0% | **10.2%** | 1.8% | **17.6%** | **17.0%** | 5.6% | - | 6.4% | 5.9% | |
| NoD (α = 1) | | 0.785 | 0.681 | 0.598 | 0.819 | 0.600 | 0.592 | 0.714 | 0.757 | 0.488 | 0.737 | 0.657 | 0.503 | 0.722 | 0.690 | 0.678 | 0.668±0.096 | |

**Table 5.** The best prediction results obtained by the CPDP approach based on *TDSelector* with Manhattan distance.

| Manhattan distance | | Ant | Xalan | Camel | Ivy | Jedit | Lucene | Poi | Synapse | Velocity | Xerces | Eclipse | Equinox | Lucene2 | Mylyn | Pde | Mean±St.d | d |
|---|---|---|---|---|---|---|---|---|---|---|---|---|---|---|---|---|---|---|
| Linear | α | 0.8 | 0.9 | 0.9 | 1.0 | 0.9 | 0.9 | 1.0 | 1.0 | 0.8 | 1.0 | 0 | 0.8 | 0.9 | 1.0 | 1.0 | | |
| | AUC | 0.804 | 0.753 | 0.599 | 0.816 | 0.689 | 0.626 | 0.695 | 0.748 | 0.500 | 0.749 | 0.773 | 0.633 | 0.692 | 0.695 | 0.668 | 0.696±0.084 | 0.187 |
| | +(%) | 1.3% | 7.0% | 0.3% | - | 7.3% | 6.3% | - | - | 7.8% | - | **11.6%** | **19.0%** | **39.7%** | - | - | 5.6% | |
| Logistic | α | 0.7 | 0.7 | 0.8 | 0.8 | 0.8 | 0.7 | 0.7 | 0.9 | 0.6 | 0.7 | 0 | 0.9 | 0.9 | 1.0 | 1.0 | | |
| | AUC | 0.799 | 0.760 | 0.607 | 0.830 | 0.674 | 0.621 | 0.735 | 0.794 | 0.520 | 0.756 | 0.773 | 0.680 | 0.559 | 0.695 | 0.668 | 0.705±0.084 | 0.249 |
| | +(%) | 0.6% | 8.0% | 1.7% | 1.7% | 5.0% | 5.4% | 5.8% | 6.1% | **12.1%** | 0.9% | **11.6%** | **27.9%** | **12.7%** | - | - | 6.9% | |
| Square root | α | 0.9 | 0.9 | 0.9 | 1.0 | 0.8 | 0.8 | 0.9 | 0.8 | 0.9 | 1.0 | 0 | 1.0 | 0 | 1.0 | 1.0 | | |
| | AUC | 0.795 | 0.755 | 0.604 | 0.816 | 0.693 | 0.627 | 0.704 | 0.750 | 0.510 | 0.749 | 0.773 | 0.532 | 0.523 | 0.695 | 0.668 | 0.680±0.1 | 0.164 |
| | +(%) | 0.1% | 7.2% | 1.2% | - | 7.9% | 6.5% | 1.3% | 0.3% | 9.9% | - | **11.6%** | - | 4.6% | - | - | 3.1% | |
| Logarithmic | α | 1.0 | 0.9 | 0.9 | 1.0 | 0.9 | 1.0 | 1.0 | 0.8 | 0.9 | 0.9 | 0 | 1.0 | 0 | 1.0 | 1.0 | | |
| | AUC | 0.794 | 0.755 | 0.603 | 0.816 | 0.664 | 0.589 | 0.695 | 0.763 | 0.524 | 0.756 | 0.773 | 0.532 | 0.523 | 0.695 | 0.668 | 0.677±0.102 | 0.116 |
| | +(%) | - | 7.2% | 1.0% | - | 3.4% | - | - | 2.0% | **12.9%** | 0.9% | **11.6%** | - | 4.6% | - | - | 2.7% | |
| Inverse cotangent | α | 1.0 | 0.9 | 0.9 | 0.9 | 0.9 | 0.8 | 0.9 | 1.0 | 0.7 | 0.8 | 0 | 1.0 | 0 | 1.0 | 1.0 | | |
| | AUC | 0.794 | 0.749 | 0.608 | 0.821 | 0.667 | 0.609 | 0.710 | 0.748 | 0.500 | 0.758 | 0.773 | 0.532 | 0.523 | 0.695 | 0.668 | 0.677±0.103 | 0.133 |
| | +(%) | - | 6.4% | 1.8% | 0.6% | 3.9% | 3.4% | 2.2% | - | 7.8% | 1.2% | **11.6%** | - | 4.6% | - | - | 2.7% | |
| NoD (α = 1) | | 0.794 | 0.704 | 0.597 | 0.816 | 0.642 | 0.589 | 0.695 | 0.748 | 0.464 | 0.749 | 0.693 | 0.532 | 0.500 | 0.695 | 0.668 | 0.659±0.105 | |

**Table 6.** Pairwise comparisons between a given combination and each of the 15 combinations in terms of the Cliff's delta (*d*) effect size.

| | Cosine similarity | | | | | Euclidean distance | | | | | Manhattan distance | | | | |
|---|---|---|---|---|---|---|---|---|---|---|---|---|---|---|---|
| | Linear | Logistic | Square root | Logarithmic | Inverse cotangent | Linear | Logistic | Square root | Logarithmic | Inverse cotangent | Linear | Logistic | Square root | Logarithmic | Inverse cotangent |
| Cosine + Linear | - | 0.018 | 0.084 | 0.000 | 0.116 | -0.049 | -0.036 | -0.004 | -0.013 | -0.009 | 0.138 | 0.049 | 0.164 | 0.178 | 0.169 |
| Euclidean + Linear | 0.049 | 0.102 | 0.111 | 0.062 | 0.164 | - | 0.036 | 0.040 | 0.058 | 0.089 | 0.209 | 0.102 | 0.249 | 0.276 | 0.244 |
| Manhattan + Logistic | -0.049 | -0.022 | 0.022 | -0.013 | 0.111 | -0.102 | -0.076 | -0.080 | -0.049 | -0.031 | 0.053 | - | 0.124 | 0.151 | 0.147 |

## 6.2 Answer to RQ2

Although the inclusion of *defects* in the selection of training data of quality is helpful for better performance of CPDP, it is worthy to note that our method completely failed in Mylyn and Pde when computing the similarity between instances in terms of Manhattan distance (see the corresponding maximum AUC values in Table 5). This implies that the success of *TDSelector* depends largely on the reasonable combination of similarity and normalization methods. Therefore, which combination of similarity and normalization is more suitable for *TDSelector*?

First of all, we analyzed the two factors (i.e. similarity and normalization) separately. For example, we evaluated the difference among Cosine similarity, Euclidean distance and Manhattan distance, regardless of any normalization method used in the experiment. The results, expressed in terms of mean and standard deviation, are shown in Table 7, where they are grouped by factors.

**Table 7.** Results of analyzing the factors similarity and normalization separately. The values of mean and standard deviation are calculated according to the best prediction results of our approach, the maximum mean value and the minimum standard deviation value for each factor appear in bold, and $d < 0$ indicates that the best similarity index or normalization method determined by mean and standard deviation performs better.

| Factor | Method | Mean | St.d | $d$ |
|---|---|---|---|---|
| Similarity | Cosine similarity | 0.704 | 0.082 | -0.133 |
| | Euclidean distance | **0.719** | **0.080** | - |
| | Manhattan distance | 0.682 | 0.098 | -0.193 |
| Normalization | Linear | 0.706 | 0.087 | -0.012 |
| | Logistic | **0.710** | **0.078** | - |
| | Square root | 0.699 | 0.091 | -0.044 |
| | Logarithmic | 0.700 | 0.086 | -0.064 |
| | Inverse cotangent | 0.696 | 0.097 | -0.056 |

If we don't take into account normalization, Euclidean distance achieves the maximum mean value 0.719 and the minimum standard deviation value 0.080 among the three similarity indexes, followed by Cosine similarity. Therefore, Euclidean distance and Cosine similarity are the first and second choices of our approach, respectively. On the other hand, if we don't take into account similarity index, the logistic normalization method seems to be the most suitable method for *TDSelector*, indicated by the maximum mean value 0.710 and the minimum standard deviation value 0.078, and it is followed by the linear normalization method. So, the logistic normalization method is the preferred way for *TDSelector* to normalize *defects*, while the linear normalization method is a possible alternative method. It is worth noting that the result is also supported by the evidence that all the Cliff's delta ($d$) effect sizes in Table 7 are negative. Then, a simple guideline for choosing appropriate similarity indexes and normalization methods for *TDSelector* from two different aspects is presented in Fig.5.

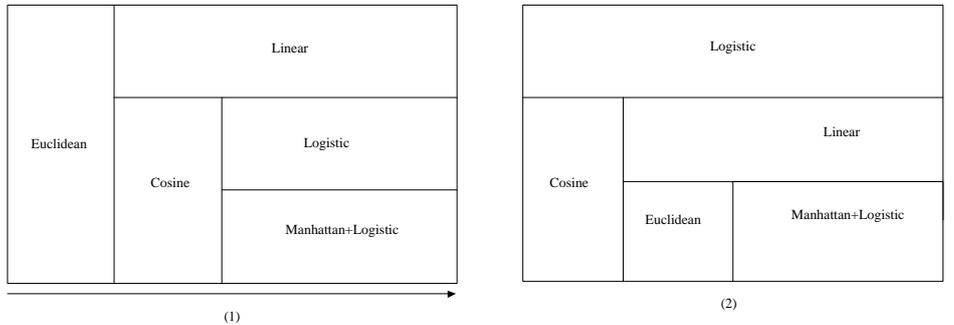

**Figure 5.** A guideline for choosing suitable similarity indexes and normalization methods from two aspects of similarity (see (1)) and normalization (see (2)). The selection priority is lowered along the direction of the arrow.

Then, we took both the two factors into account. According to the results in Tables 3, 4 and 5 grouped by different similarity indexes, *TDSelector* can obtain the best result 0.719±0.080, 0.711±0.070, and 0.705±0.084 when using "Euclidean + Linear" (short for Euclidean distance + Linear Normalization), "Cosine + Logistic" (short for Cosine similarity + Logistic Normalization), and "Manhattan + Logistic" (short for Manhattan distance + Logistic Normalization), respectively. We also calculated the value of the Cliff's delta ($d$) effect size for each two combinations under discussion. As shown in Table 6, according to the largest number of positive $d$ values in this table, the combination of Euclidean distance and the linear normalization method can still outperform the other 14 combinations.

In brief, the best combination of similarity and normalization for *TDSelector* is "Euclidean + Linear", and an alternative combination is "Cosine + Linear" if you choose Cosine similarity as the similarity index (see Table 6 and Fig.5).

### 6.3 Answer to RQ3

A comparison between our approach and two baseline methods (i.e. baseline1 and baseline2) across the 15 test sets is presented in Table 8, where the second and third columns represent the maximum AUC values achieved by the two baseline methods, and the fourth and fifth columns indicate their corresponding growth rates of AUC value obtained by the TDSelector that was built using the combination of Euclidean distance and Linear Normalization.

It is obvious that our approach is, on average, better than the two baseline methods, indicated by the average growth rates of AUC value (i.e. 10.6% and 4.3%) across the 15 test sets. The TDSelector performs better than baseline1 in 14 out of 15 data sets, and it has an advantage over baseline2 in 10 out of 15 data sets. In particular, compared with baseline1 and baseline2, the highest growth rates of AUC value of our approach reach up to 65.2% and 64.7%, respectively, for Velocity. Besides, the negative $d$ values in this table also indicate that our approach outperforms the baseline methods from the perspective of distribution, though we have to admit that the Cliff's delta effect size value of 0.009 is too small to be of interest in a particular application.

**Table 8.** A comparison between our approach and two baseline methods for the data sets from PROMISE and AEEEM. The comparison is conducted based on the best prediction results of all the three methods in question, and $d < 0$ indicates that our approach performs better than the corresponding baseline method.

| Test set | Baseline1 | Baseline2 | Euclidean + Linear | | $d$ |
|---|---|---|---|---|---|
| Ant | 0.785 | 0.803 | 1.3% | -1.0% | |
| Xalan | 0.657 | 0.675 | 10.7% | 7.7% | |
| Camel | 0.595 | 0.624 | 0.5% | -4.2% | |
| Ivy | 0.789 | 0.802 | 4.7% | 3.0% | Baseline1 vs. TDSelecotr: -0.409 |
| Jedit | 0.694 | 0.782 | 14.3% | 1.4% | |
| Lucene | 0.608 | 0.701 | -0.8% | -14.0% | |
| Poi | 0.691 | 0.789 | 3.3% | -9.5% | |
| Synapse | 0.740 | 0.748 | 2.3% | 1.2% | |
| Velocity | 0.330 | 0.331 | 65.2% | 64.7% | |
| Xerces | 0.714 | 0.753 | 8.5% | 2.9% | |
| Eclipse | 0.706 | 0.744 | 10.2% | 4.6% | |
| Equinox | 0.587 | 0.720 | 23.1% | 0.3% | Baseline2 vs. TDSelector: -0.009 |
| Lucene2 | 0.705 | 0.724 | 2.5% | -0.2% | |
| Mylyn | 0.631 | 0.646 | 9.3% | 6.8% | |
| Pde | 0.678 | 0.737 | 10.4% | 1.5% | |
| Avg. | 0.663 | 0.705 | 10.6% | 4.3% | |

In summary, since the *TDSelector*-based defect predictor outperforms those based on the two state-of-the-art CPDP methods, our approach is beneficial for training data selection and can further improve the performance of CPDP models.

## 7 Discussion

### 7.1 Impact of Top-*k* on Prediction Results

The parameter *k* determines the number of the nearest training instances of each test instance. Since *k* was set to 10 in our experiments, here we discuss the impact of *k* on prediction results of our approach as its value is changed from 1 to 10 with a step value of 1. As shown in Fig.6, for the three combinations in question, selecting the *k*-nearest training instances (e.g. $k \leq 5$) for each test instance in the 10 test sets from PROMISE, however, doesn't lead to better prediction results, because their best results are obtained when *k* is equal to 10.

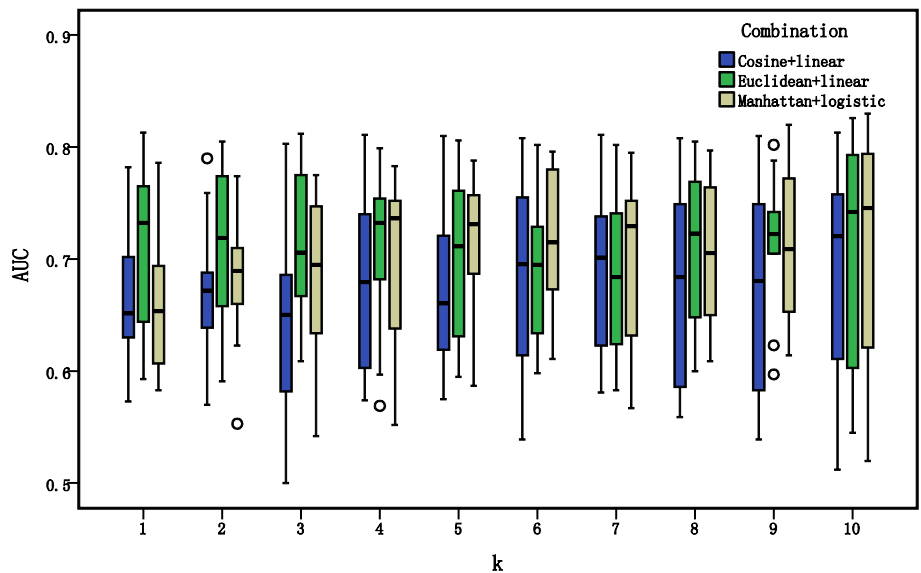

**Figure 6.** The impact of *k* on prediction results for the 10 test sets from PROMISE

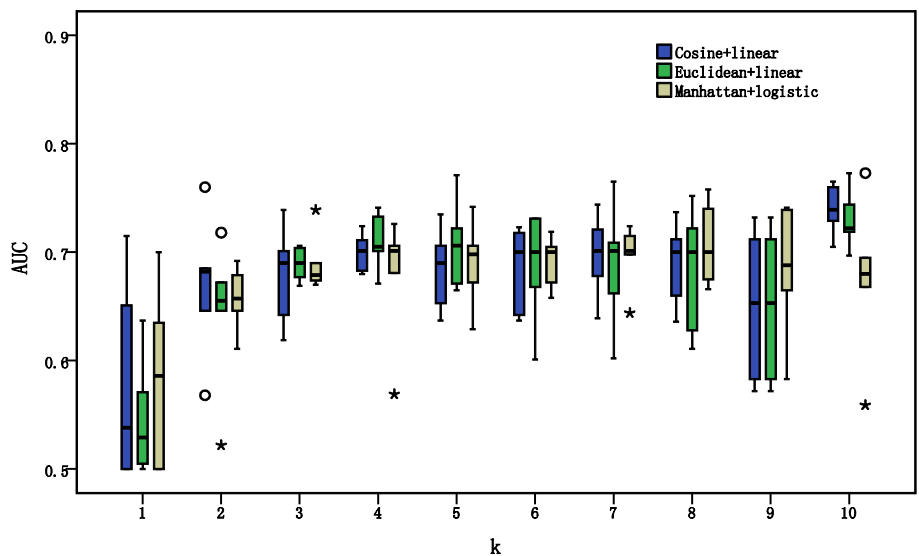

**Figure 7.** The impact of *k* on prediction results for the 5 test sets from AEEEM

Interestingly, for the combinations of "Euclidean + Linear" and "Cosine + Linear", a similar trend of AUC value changes is visible in Fig.7. For the five test sets from AEEEM, they achieve stable prediction results when $k$ ranges from four to eight, and then reach peak performance when $k$ is equal to 10. The combination of "Manhattan + Logistic", by contrast, achieves the best result as $k$ is set to seven. Even so, its best result is still worse than those of the other two combinations.

## 7.2 Selecting instances with more bugs directly as training data

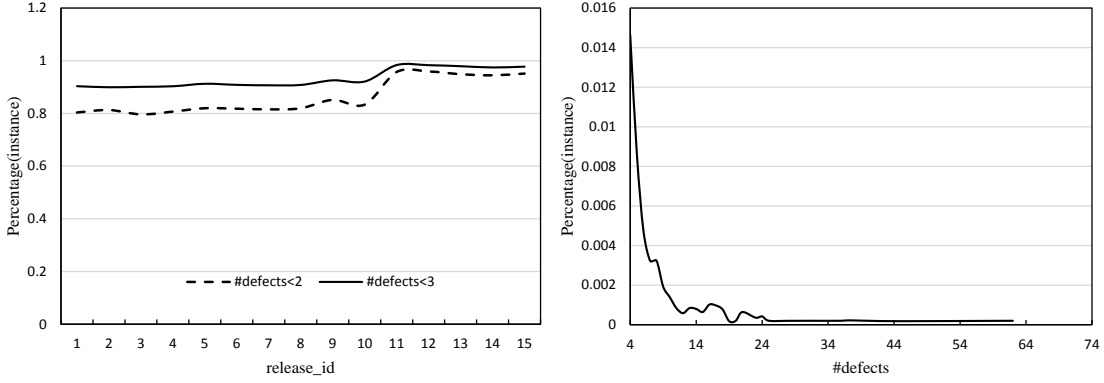

**Figure 8.** Percentage of defective instances with different number of bugs. The left plot is shown from the viewpoint of a single data set (release), while the right plot is shown from the viewpoint of the whole data set used in our experiments.

Our experimental results have validated the impact of *defects* on the selection of training data of quality in terms of AUC, and we also want to know whether the direct selection of defective instances with more bugs as training instances, which simplifies the selection process and reduces computation cost, would achieve better prediction performance. The result of this question is of particular concern for developers in practice.

According to the left plot in Fig.8, for the 15 releases, most of them contain instances with no more than two bugs. On the other hand, the ratio of the instances that have more than three defects to the total instances is less than 1.40% (see the right plot in Fig.8). Therefore, we built a new TDSelector based on the number of bugs in each instance, which is referred to as TDSelector-3. That is to say, those defective instances that have at least three bugs were chosen directly from an initial TDS as training data, while the remaining instances in the TDS were selected in the light of Eq.(2). All instances from the two parts then form the final TDS after removing redundant ones.

Fig.9 shows that the results of the two methods differ from data set to data set. For Ivy and Xerces collected from PROMISE, TDSelector outperforms TDSelector-3 in all the three scenarios, but only slightly. On the contrary, for Lucene and Velocity from PROMISE, the incremental AUC values obtained by using TDSelector-3 with "Cosine + Linear" reach up to 0.109 and 0.235, respectively. As shown in Fig.9, on average, TDSelector-3 performs better than the corresponding TDSelector, and the average AUC values for "Cosine + Linear", "Euclidean + Linear" and "Manhattan + Logistic" are improved by up to 3.26%, 2.57% and 1.42%, respectively. Therefore, the direct selection of defective instances that contain quite a few bugs can, on the whole, further improve the performance of the predictor trained by our approach. In other words, those valuable defective instances can be screened out quickly according to a threshold for the number of bugs in each training instance (viz. three in this paper) at the first stage. Our approach is then able to be applied to the remaining TDS. Note that the automatic optimization method for such a threshold for *TDSelector* will be investigated in our future work.

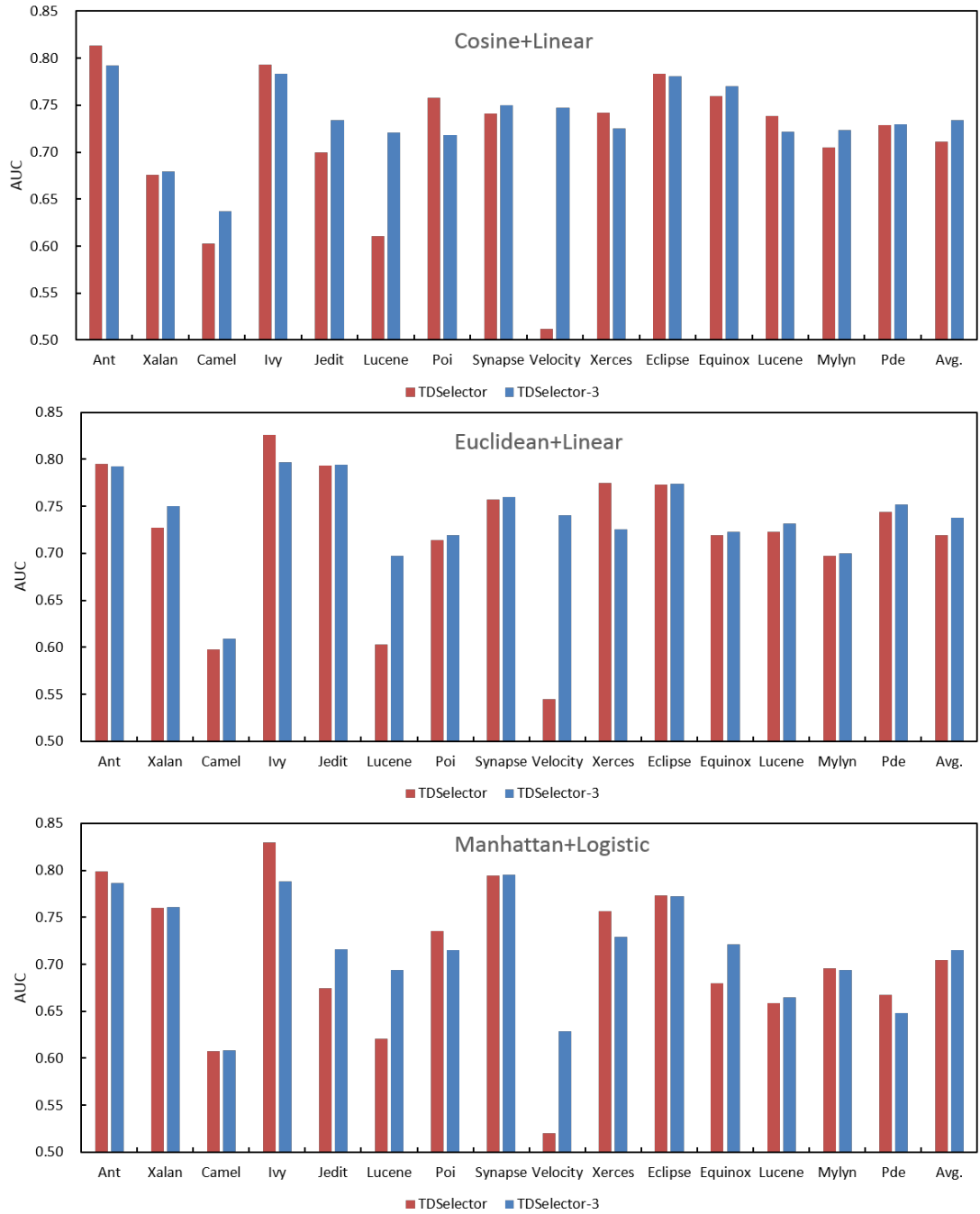

**Figure 9.** A comparison of prediction performance between TDSelector-3 and the corresponding TDSelector. The last column in each of the three plots represents the average AUC value.

## 7.3 Threats to Validity

In this study, we obtained several interesting results, but potential threats to the validity of our work still remain.

Threats to *internal validity* concern any confounding factor that may affect our results. First, the raw data used in this paper were normalized by using the Z-score method, while the baseline method TCA+ provides four normalization methods [35]. Second, unlike TCA+, *TDSelector* does not introduce any feature selection method to process software metrics. Third, the weighting factor

*α* changes with a step size 0.1, when Algorithm 1 calculates the maximum value of AUC. There is no doubt that a smaller step size will result in greater calculation time. Fourth, we trained only one type of defect predictors based on the default parameter settings configured by the tool Weka, because LR has been widely used in previous studies. Hence, we are indeed aware that the results of our study would change if we use different settings of the above three factors.

Threats to *statistical conclusion validity* focus on whether conclusions about the relationship among variables based on the experimental data are correct or reasonable [45]. In addition to mean value and standard deviation, in this paper we also utilized the Cliff's delta effect size instead of hypothetical test methods such as the Kruskal–Wallis H test [46] to compare the results of different methods, because there are only 15 data sets collected from PROMISE and AEEEM. According to the criteria that were initially suggested by Cohen and expanded by Sawilowsky [47], nearly all of the effect size values in this paper belong to *small* ($0.2 \leq d < 0.5$) and *very small* ($0.01 \leq d < 0.2$). This indicates that there is no significant difference in AUC value between different methods in question, though our method performs better than those baseline methods in terms of mean value and standard deviation.

Threats to *external validity* emphasize the generalization of the obtained results. First, the selection of experimental data sets—in addition to AEEEM and PROMISE—is the main threat to validate the results of our study. All the 14 projects used in this paper are written in Java and from the Apache Software Foundation and the Eclipse Foundation. Although our experiments can be repeated with more open-source projects written in other programming languages and developed with different software metrics, the empirical results may be different from our main conclusions. Second, we utilized only three similarity indexes and five normalization methods when calculating the score of each candidate training instance. Therefore, the generalizability of our method for other similarity indexes (such as Pearson Correlation Coefficient and Mahalanobis distance [48]) and normalization methods has yet to be tested. Third, to compare our method with TCA+, defect predictors used in this paper were built using LR, implying that the generalizability of our method for other classification algorithms remains unclear.

## 8 Conclusion and Future Work

This study aims to train better defect predictors by selecting the most appropriate training data from those defect data sets available on the Internet, so as to improve the performance of cross-project defect predictions. In summary, the study has been conducted on 14 open source projects and consists of (1) an empirical validation on the usability of the number of defects that an instance includes for training data selection, (2) an in-depth analysis of our method *TDSelector* with regard to similarity and normalization, and (3) a comparison between our proposed method and the benchmark methods.

Compared with those similar previous studies, the results of this study indicate that the inclusion of *defects* does improve the performance of CPDP predictors. With a rational balance between the similarity of test instances with training instances and *defects*, *TDSelector* can effectively select appropriate training instances, so that *TDSelector*-based defect predictors, built by using LR, achieve better prediction performance in terms of AUC. More specifically, the combination of Euclidean distance and Linear normalization is the preferred way for *TDSelector*. In addition, our results also demonstrate the effectiveness of the proposed method according to a comparison with the baseline methods in the context of M2O in CPDP scenarios. Hence, we believe that our approach can be helpful for developers when they are required to build suitable predictors quickly for their new projects, because one of our interesting findings is that those candidate instances with more bugs can be chosen directly as training instances.

Our future work mainly includes two aspects. On the one hand, we plan to validate the generalizability of our study with more defect data from projects written in different languages. On the other hand, we will focus on more effective hybrid methods based on different selection strategies such as feature selection techniques [28]. Last but not least, we also plan to discuss the possibility of considering not only the number of defects but also time variables for training data selection (such as bug-fixing time).

## Acknowledgement

We greatly appreciate Dr. Jaechang Nam and Dr. Sinno Jialin Pan, the authors of the reference [35], for providing us the TCA source program and friendly teaching us how to use it.

This work was supported by the National Basic Research Program of China (No. 2014CB340404), the National Key Research and Development Program of China under the Grant No. 2016YFB0800400, the National Natural Science Foundation of China (Nos. 61272111, 61273216, 61572371 and 61672387), and the Wuhan Yellow Crane Talents Program for Modern Services Industry.